# Strong light-induced negative optical pressure arising from the kinetic energy of conduction electrons in plasmonic cavities

H. Liu[1,2*], Jack Ng[2], S. B. Wang[2], Z. F. Lin[2, 3], Z. H. Hang[2], C. T. Chan[2,*] and S. N. Zhu[1]

[1]*Department of Physics, National Laboratory of Solid State Microstructures, Nanjing University, Nanjing 210093, People's Republic of China*

[2]*Department of Physics and Nano Science and Technology Program , Hong Kong University of Science and Technology, Clearwater Bay, Hong Kong, China*

[3]*Department of Physics, Fudan University, Shanghai, China*

**Corresponding Authors: liuhui@nju.edu.cn; phchan@ust.hk**

## Abstract

We found that very strong negative optical pressure can be induced in plasmonic cavities by LC resonance. This interesting effect could be described qualitatively by a Lagrangian model which shows that the negative optical pressure is driven by the internal inductance and the kinetic energy of the conduction electrons. If the metal is replaced by perfect conductors, the optical pressure becomes much smaller and positive.

The fact that electromagnetic (EM) radiation exerts pressure on any surface was deduced theoretically by Maxwell in 1871, and demonstrated experimentally by Lebedev in 1900 [1] and by Nichols and Hull in 1903 [2]. Recently, there is an increasing interest in optical forces acting on resonance cavities [3-8], partly due to the emergence of the field of cavity optomechanics [7-11]. One may anticipate that the radiation field confined by a cavity should serve to expand the cavity as the photons bounce back and forth between the cavity walls. This is indeed the case for a large class of resonant cavities, such as closed metallic rectangular cavities made up of perfect electric conductor (PEC) walls, whispering gallery mode cavities, and Fabry-Perot cavities. We will show that a very simple system can display light-induced resonance that is not only large in magnitude, but also attractive rather than repulsive. This will provide a new platform to realize giant optical forces, in addition to micro/nano-systems such as photonic crystals [12, 13], plasmonic structures [14-17], micro-cavities and waveguides [18-20].

We propose that instead of the usual positive pressure, one can induce a strong negative pressure between the walls of an open metallic cavity, by utilizing the kinetic energy of the electrons inside the metallic cavity walls. Consider a planar structure consisting of a square metal patch placed above another bigger square metal slab, as illustrated in Fig. 1. The thickness of the patch and slab is t=10nm. The size of slab is taken to be five times of the patch, and the results are qualitatively the same as long as the slab is a few times the size than the patch. The distance between the patch and the slab is D. Metamaterials [21-26] exhibit artificial resonant responses, and such responses can be realized in the visible or infrared frequency range based on LC resonance effect [27-29]. Here, the patch and slab forms an LC resonance cavity [30]. We propose a simple Lagrangian model to describe qualitatively the resonance property of the equivalent LC circuit for this structure [31-34]. The circuit's Lagrangian could be expressed as [35].

$$\mathcal{L} = \frac{L\cdot\dot{Q}^2}{2} - \frac{Q^2}{2\cdot C} \qquad (1)$$

Here, $L = L_m + L_e$ and $C = C_m + C_e$ . $L_m = \mu_0 D/2$ and $C_m$ are the external inductance and capacitance, $L_e = 1/\left(\omega_p^2 \varepsilon_0 t\right)$ and $C_e$ are the internal inductance and capacitance [36-38], Q is the net charge in the capacitor, $L_m\dot{Q}^2/2$ is the magnetic field energy stored between the two patches, $U_{kinetic} = L_e\dot{Q}^2/2$ is the kinetic energy of the electrons due to the induced current inside the metal. We will ignore the internal capacitance which is much smaller than external capacitor, i.e. $C \approx C_m$ . Then $U_{electric} = Q^2/(2C_m)$ is the potential energy stored in the cavity. For a finite sized patch, the edge effect due to field leakage on both ends of the patch has to be considered, and $C_m$ is modeled as $C_m = \varepsilon_0 \frac{2}{\pi^2}\frac{A^2}{D}\cdot\left(1+\alpha\frac{D}{A}\right)$ [35], where the empirical coefficient $\alpha$ gives the magnitude of edge effect.

In the presence of Ohmic dissipation and an external driving field, the Euler-Lagrange equation can be written as $\frac{d}{dt}\left(\frac{\partial \mathcal{L}}{\partial \dot{Q}}\right)-\frac{\partial \mathcal{L}}{\partial Q}=-\frac{\partial \mathcal{R}}{\partial \dot{Q}}+\text{e.m.f.}$, where $\mathcal{R}=R_{eff}\cdot\dot{Q}^2/2$ ($R_{eff}$ is the effective resistance in the system) and the e.m.f. due to external field $E_0$ is $\text{emf}=\left(-\oint E\cdot d\ell\right)=-\left(\ell_{eff}^{(1)}\cdot E_0-\ell_{eff}^{(2)}\cdot E_0\cdot e^{ikD}\right)$ with $\ell_{eff}^{(1)}$ and $\ell_{eff}^{(2)}$ being the effective length of the dipole on the patch and the slab respectively and the factor $\exp(ik\cdot D)=\exp\left(i\frac{\omega}{c}\cdot D\right)$ represents the retardation effect. To simplify expressions, we introduce the notation $\tau=\ell_{eff}^{(2)}/\ell_{eff}^{(1)}$ [35]. Solving the Euler-Lagrange equation gives the resonance frequency of the cavity $\omega_0=\frac{1}{\sqrt{(L_e+L_m)\cdot C_m}}=\frac{1}{\sqrt{\left(\frac{1}{\varepsilon_0\omega_p^2 t}+\frac{\mu_0 D}{2}\right)\cdot\varepsilon_0\frac{2}{\pi^2}\frac{A^2}{D}\cdot\left(1+\alpha\cdot\frac{D}{A}\right)}}$. The optical force exerted on the patches can be calculated as a generalized force corresponding to the coordinate D [35]:

$$\mathcal{F}=\frac{\partial \mathcal{L}}{\partial D}=\frac{1}{2}\left(\frac{\partial L_m}{\partial D}+\frac{\partial L_e}{\partial D}\right)\dot{Q}^2+\frac{1}{2}\frac{Q^2}{C_m^2}\left(\frac{\partial C_m}{\partial D}\right). \tag{2}$$

As the internal inductance does not depend on the gap distance, we have $\partial L_e/\partial D=0$ [36-38]. Approximating our system by a parallel plate transmission line, we have, $L_m\sim D\Rightarrow \partial L_m/\partial D=L_m/D$. We also have $\frac{\partial C_m}{\partial D}=-\frac{C_m}{D}\cdot\frac{A}{A+\alpha\cdot D}$. At the resonance frequency $\omega_0$, if time averaging is performed for one oscillation period, one could obtain $L\left\langle\dot{Q}^2\right\rangle_t=(L_m+L_e)\cdot\omega_0^2\cdot\left\langle Q^2\right\rangle_t=\left\langle Q^2\right\rangle_t/C_m$. Substituting these into Eq. (2), one obtains

$$\left\langle\mathcal{F}\right\rangle_t=-\frac{1}{2}\frac{L_e\left\langle\dot{Q}^2\right\rangle_t}{D}+\frac{1}{2}\frac{\left\langle Q^2\right\rangle_t}{C_m}\frac{1}{D}\left(\frac{\alpha D}{A+\alpha D}\right). \tag{3}$$

Here, $\langle\cdot\rangle_t$ denotes the time averaging. We can see that the total optical force given in Eq. (3) includes a term $-\frac{1}{2}\frac{L_e\left\langle\dot{Q}^2\right\rangle_t}{D}$ which is the attractive force from kinetic energy of electron and a term $\frac{1}{2}\frac{\left\langle Q^2\right\rangle_t}{C_m}\frac{1}{D}\left(\frac{\alpha D}{A+\alpha D}\right)$ which is the repulsive force from E field edge effect. If the edge effect is small, the optical force will be dominated by an attractive force $-\frac{1}{2}\frac{L_e\left\langle\dot{Q}^2\right\rangle_t}{D}$ derived from kinetic energy of the electron in the induced current. At the resonance frequency $\omega_0=\frac{1}{\sqrt{(L_m+L_e)\cdot C_m}}$ ( $k_0=\frac{\omega_0}{c}$ ), $\frac{1}{2}\frac{\left\langle Q^2\right\rangle_t}{C_m}=\frac{1}{2}\frac{(L_m+L_e)\left\langle\dot{Q}^2\right\rangle_t}{D}$, $\left\langle\dot{Q}^2\right\rangle_t=\left(\frac{\ell_{eff}^{(1)}}{R_{eff}}\right)^2\cdot\left[\frac{1+\tau^2}{2}-\tau\cdot\cos(k_0\cdot D)\right]\cdot E_0^2$. The effective length of patch could be approximately taken as $\ell_{eff}^{(1)}\simeq A$, and the area of patch is $S=A^2\simeq\left(\ell_{eff}^{(1)}\right)^2$. The optical pressure normalized by the incident intensity $I_0=\frac{1}{2}\varepsilon_0\cdot c\cdot E_0^2$ could be obtained as [35] $P=\frac{\eta}{D}\cdot\left[\frac{1+\tau^2}{2}-\tau\cdot\cos(k_0\cdot D)\right]\cdot\left(-1+(1+\beta D)\cdot\frac{\alpha D}{A+\alpha D}\right)$, where $\eta=\frac{t}{c}\cdot\frac{\omega_p^2}{\gamma_m^2}$, $\beta=\frac{t}{2}\cdot\frac{\omega_p^2}{c^2}$. Equation (3) suggests that the optical pressure between the the patch and the slab is induced by the LC resonance is determined by competition between negative pressure from the kinetic energy of electrons and positive pressure from the edge effect of E field. In the following numerical simulations, we will show that the negative pressure from kinetic energy is much stronger than the positive pressure from edge effect. Then the total optical pressure P is negative.

We shall now proceed by using numerical simulation with a commercial software package CST Microwave to show that our Lagrangian model can capture the essence of physics and a giant negative pressure can indeed be induced by external light. In the simulations, the metal permittivity gold is taken to be of the Drude form $\varepsilon(\omega)=1-\frac{\omega_p^2}{(\omega^2+i\gamma_m\cdot\omega)}$, with $\omega_p=1.37\times10^{16}s^{-1}$ and $\gamma_m=12.08\times10^{13}s^{-1}$ for gold. These characteristic frequencies are fitted from experimental data [39]. For the structure depicted in Fig. 1, the parameters are chosen as A=200nm, D=30nm and t=10nm in the simulations. Open boundary conditions are used in all three directions. The incident plane EM propagates upward along the y direction. The E field of incident wave is set as $E_0=1.0V/m$. When the incident frequency is swept from 150 to 350 THz, the frequency dependence of E field between two patches is plotted in Fig. 2 (a). One LC resonance mode is excited at 255THz. The profiles of the electric and magnetic field at the LC resonance mode are shown in fig. 2 (c)-(d). We see that the field strongly localizes in the space between the patch and the slab at this resonance frequency.

After obtaining the EM fields, the time-averaged optical force $\langle\mathcal{F}\rangle_t$ between the patch and slab can then be calculated rigorously using the Maxwell's stress tensor via a surface integral $\langle\mathcal{F}\rangle_t=\oiint_S\langle T\rangle_t\cdot ds$, where $T_{\alpha\beta}=\varepsilon_0\left(E_\alpha E_\beta-E\cdot E\delta_{\alpha\beta}/2\right)+\mu_0\left(H_\alpha H_\beta-H\cdot H\delta_{\alpha\beta}/2\right)$ and S is the integration surface enclosing one patch. Figure 2(b) shows the calculated optical force per unit area acting on the surface of patch as a function of frequency. At the resonance frequency, a strong negative (attractive) pressure on the wall of cavity could be obtained (about -698 Pa/(mW/μm$^2$)). For an infinite perfectly reflecting plate, the positive optical pressure exerted by a plane wave with the same frequency is only about 10 Pa/(mW/μm$^2$). The negative optical pressure in the plasmonic cavity is much stronger than the usual optical pressure. Such a strong negative optical pressure might be able to mechanically deform the structure, leading to optomechanical coupling of the photon and phonon modes [7-11]. The direction of the force is opposite to the propagating direction of the incident light [1, 2].

The dependence of the optical pressure on the distance D was numerically simulated and plotted in Fig. 3 (a). For small D ($D\ll A$), as the edge effect term in equation (3) can be neglected, P is dominated by the kinetic energy term, which carries an explicit 1/D dependence. Consequently, the log of ($-P$) should scale linearly with log(D) for small D. The log-log relationship between optical pressure P and distance D is given as black square dots in Fig. 3 (a). As shown in Fig. 3 (b), for the small distance D below 50nm, $\log(-P)$ is indeed inversely proportional to $\log(D)$. For D greater than 50nm, the edge effect term in Eq. (3) cannot be ignored and its positive pressure will counteract part of the negative pressure from the kinetic energy term. As a result, the total optical pressure is no longer inversely proportional to D and the calculated $\log(-P)$ deviates from a straight line progressively as D increases, as shown in Fig. 3 (a). In our CST simulations, the dependence of the optical pressure on the parameter A was also investigated and plotted in Fig. 3 (b). There is an optimal patch size that gives a maximum force, and the optical pressure decreases for smaller and bigger patches. We show in the Appendix [34] that the salient features of the numerical calculations can be reproduced with the Lagrangian model. In particular, the Lagrangian model gives a simple picture to show that the optical pressure in plasmonic cavity is mainly driven by the near-field localized LC resonance coupling and the kinetic energy inside the plasmonic plates is the key behind the strong negative pressure.

To further clarify the role of the kinetic energy of electrons, we consider the same structure illustrated in Fig. 1 with A =200nm and D=30nm, but with all metals replaced by perfect conductors (PEC). Since the EM field does not penetrate into the PEC, there is no contribution from the kinetic energy of the electrons. According to our model, the optical pressure only comes from the edge effect term in Eq. (3), which should be repulsive and small as A is much bigger than D. Indeed, the numerical calculated optical force for the PEC structure is only 187 Pa/(mW/μm$^2$), much smaller than the attractive force of -698 Pa/(mW/μm$^2$) in the plasmonic structure (see table 1). Moreover, the optical pressure for the PEC structure is positive, contrary to the negative pressure in plasmonic systems. While the numerical calculations are fully electrodynamical, we can understand the results qualitatively by considering the quasi-static limit, as we have a subwavelength system. In that limit, the optical force can be separated into a Coulomb electric force $\langle \mathcal{F}_e \rangle_t = \oiint \langle T^e_{\alpha\beta} \rangle_t \cdot ds$ and an Ampere magnetic force $\langle \mathcal{F}_m \rangle_t = \oiint \langle T^m_{\alpha\beta} \rangle_t \cdot ds$, where $T^e_{\alpha\beta} = \varepsilon_0 \left( E_\alpha E_\beta - E \cdot E \delta_{\alpha\beta} / 2 \right)$ and $T^m_{\alpha\beta} = \mu_0 \left( H_\alpha H_\beta - H \cdot H \delta_{\alpha\beta} / 2 \right)$. Since the former is induced by charges and the latter by current, we can write $\langle \mathcal{F}_e \rangle_t = \oiint \langle T^e_{\alpha\beta} \rangle_t \cdot ds = -\frac{\langle Q^2 \rangle_t}{C} \frac{1}{D}$ and $\langle \mathcal{F}_m \rangle_t = \oiint \langle T^m_{\alpha\beta} \rangle_t \cdot ds = \frac{L_m}{D} \langle \dot{Q}^2 \rangle_t$. As the size of the patch is finite, the E field at the edge of the patch cannot be completely confined under the patch. Table 1 shows that this edge effect of E field makes the attractive Coulomb force (-299.5 Pa/(mW/μm$^2$) smaller than the repulsive Ampere force (487.3 Pa/(mW/μm$^2$), and the difference of 187.9 Pa/(mW/μm$^2$) is the residual positive pressure due to edge effect as described in the Lagrangian model. However, the situation is entirely different for the plasmonic cavity. In the structure made of Drude metal and at frequencies in which the field penetrates the metal, most of the inductance is manifested as the kinetic energy of the electrons and only a small part in the magnetic field. The electric field energy becomes stronger than the magnetic field energy, and this difference induces strong optical forces in the plasmonic structure. In our simulated data given in Table 1, the negative optical pressure from the electric field (-849 Pa/(mW/μm$^2$)) is much larger than the positive optical pressure from the magnetic field (151 Pa/(mW/μm$^2$)). These results agree with our theoretical model quite well.

In the above discussion, we focused on the optical pressure in a single plasmonic cavity. For a periodic array of many such plasmonic cavities, the total optical force could be seen as sum of optical force from many such plasmonic resonators if the coupling interaction between them is small, and coherent coupling may further enhance the effect. This could produce a substantial optical force in an extended system. Not only is this optical force strong, the phenomenon is expected to be very robust. We can see from Fig. 2(b) that the quality factor is relatively low, at least when compared with, for example, whispering gallery mode resonators. In principle, one could also obtain strong optical forces between two objects if high-fidelity resonances (such as whispering-gallery mode) are excited, but the frequency must be very precise in those "photonic molecules" and would be much more difficult to realize than the present configuration which employs plasmonic resonances. We note that in ordinary electromagnetic cavities, such as Fabry-Perot cavities or micro-disk resonators, the resonance force is repulsive. In our plasmonic system, the penetration of the field into the metal leads to a giant attractive optical force.

In summary, we designed a plasmonic cavity system comprising a patch and a slab. We find that the kinetic energy of conduction electrons plays a key role in inducing a strong negative optical pressure. A Lagrangian model is proposed to describe the salient features. The mechanism and theoretical model reported in this paper could have potential applications in many other subwavelength optomechanical plasmonic structures.

The research of H.L. was financially supported by the National Natural Science Foundation of China (No.10704036, No.10874081, No.60907009, No.10904012, No.10974090 and No. 60990320), and by the National Key Projects for Basic Researches of China (No. 2006CB921804, No. 2009CB930501 and No. 2010CB630703). The work in Hong Kong is supported by RGC Hong Kong through grant 600308 and by Nano Science and Technology Program of HKUST. Z.F.L. was supported by NSFC (10774028), CNKBRSF (2006CB921706), PCSIRT, MOE of China (B06011), and Shanghai Science and Technology Commission. Computation resources are supported by Hong Kong Shun Hing Education and Charity Fund.

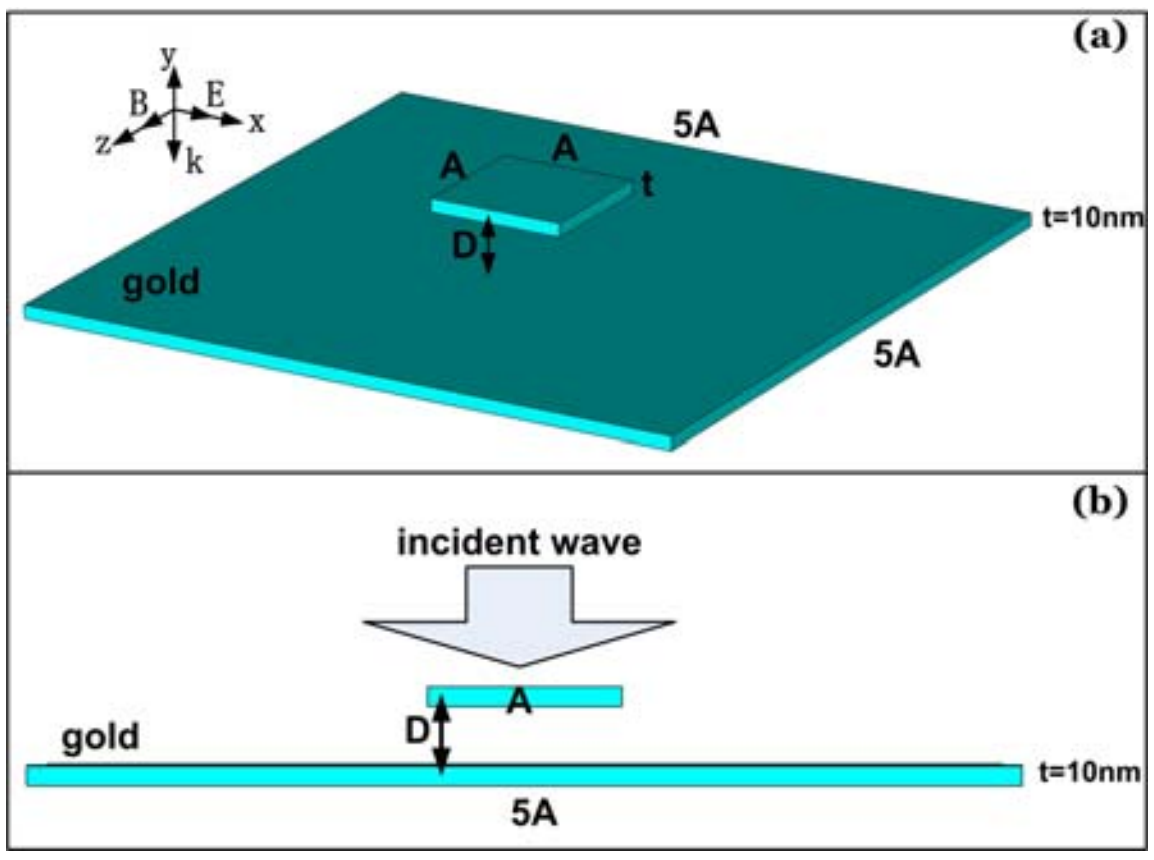

Fig.1 Schematic of nanocavity with two gold patches.

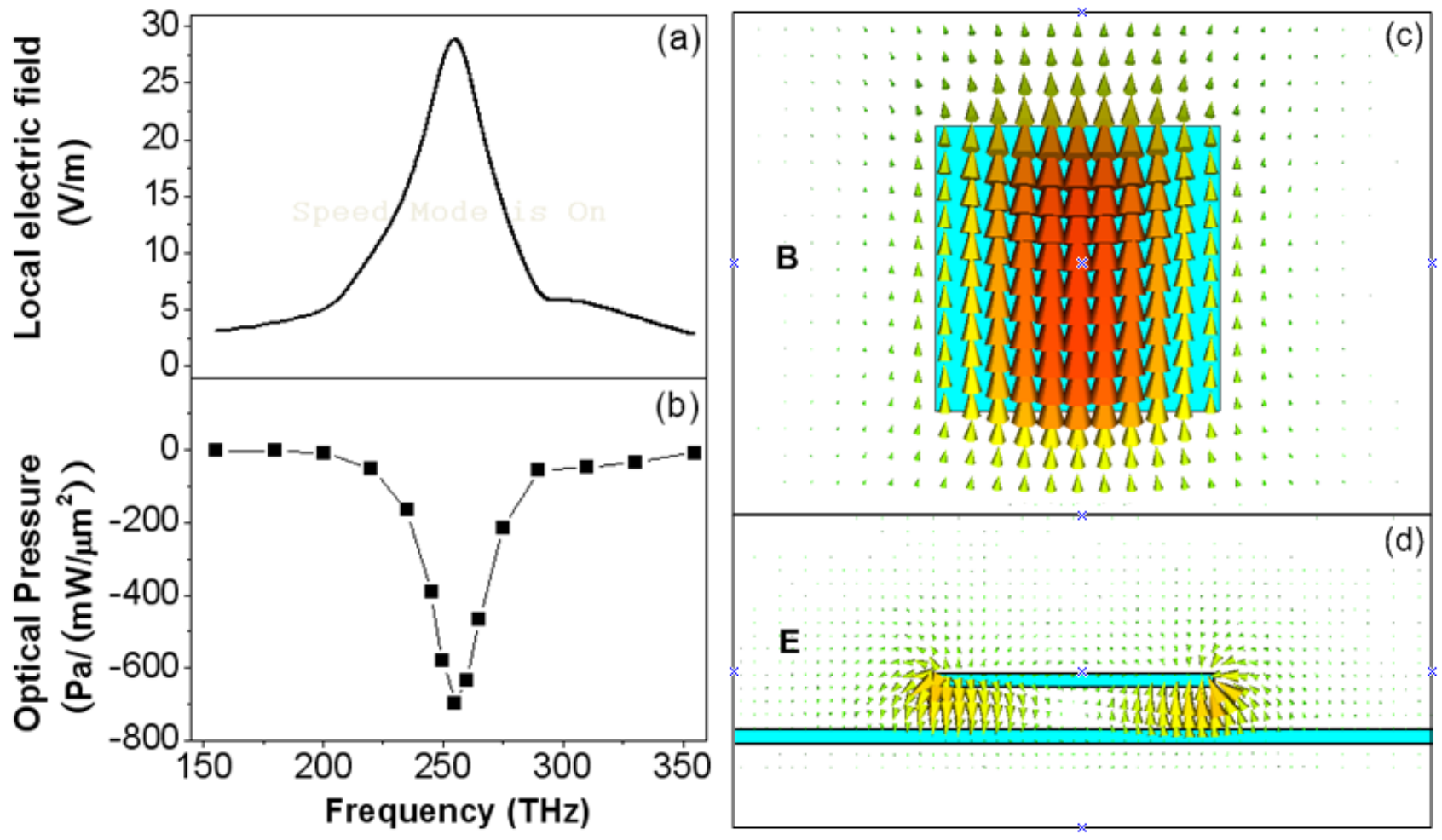

Fig.2 The frequency dependence of (a) local electric field between two patches and (b) optical pressure between two patches (with A=200nm, D=30nm); (c) Magnetic field (on y-cut middle layer) and electric field (on z-cut middle layer) at the resonance frequency $\omega = 255\text{THz}$ .

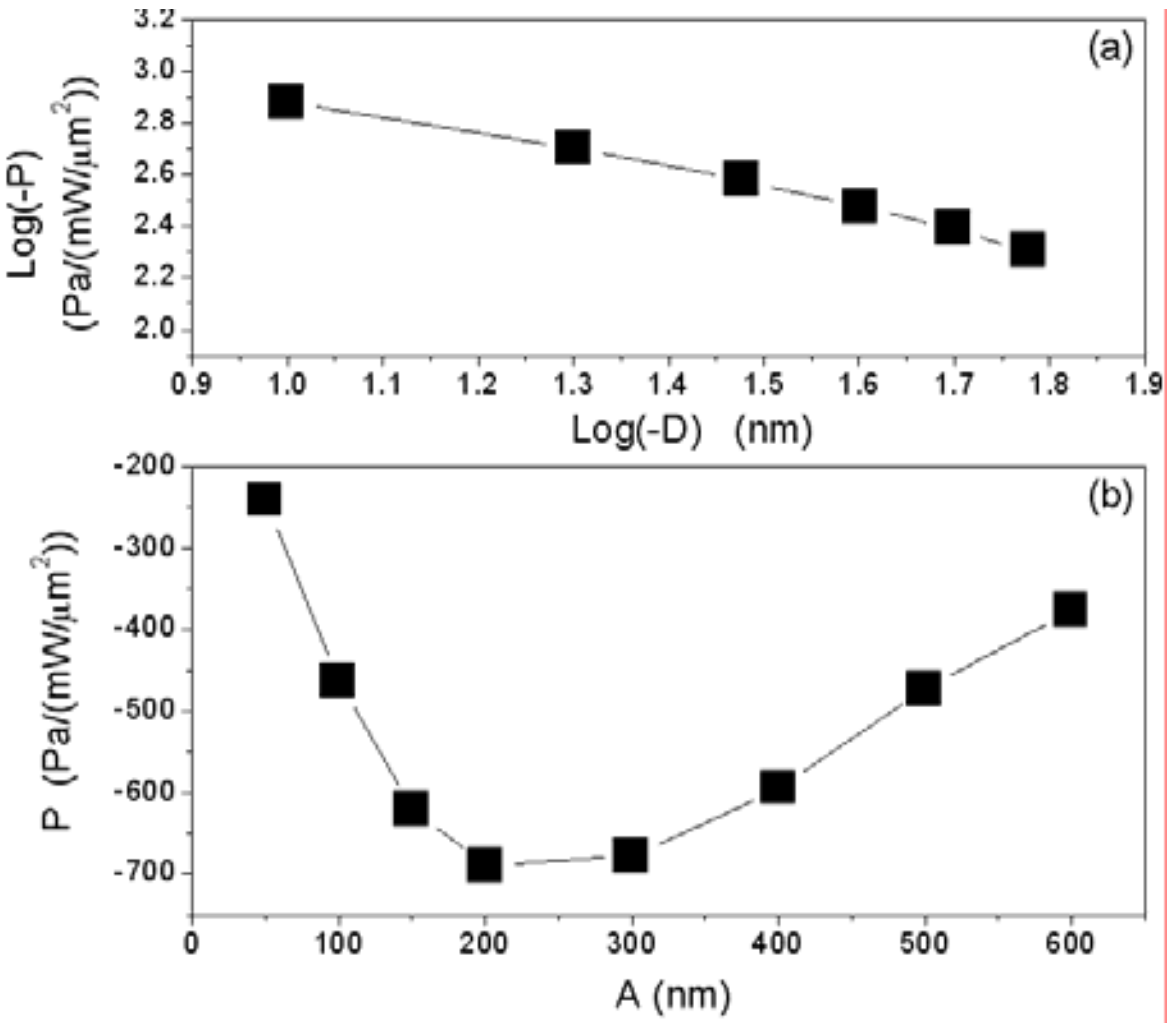

Figure 3 (a) The log-log relationship between optical pressure P and distance D (with A=600nm); (b) the dependence of optical pressure P on the size parameter A (with D=30nm).

Table 1 Numerically calculated optical forces for the plasmonic and PEC structure (with A =200nm, D=30nm) at resonance frequencies.

| Material | Optical force calculated from magnetic field [Pa/(mW/μm$^2$)] $\langle \mathcal{F}_m \rangle_t = \oiint \langle T^m_{\alpha\beta} \rangle_t \cdot ds$ | Optical force calculated from electric field [Pa/(mW/μm$^2$)] $\langle \mathcal{F}_e \rangle_t = \oiint \langle T^e_{\alpha\beta} \rangle_t \cdot ds$ | Total optical pressure [Pa/(mW/μm$^2$)] $\langle \mathcal{F} \rangle_t = \oiint \left[ \langle T^e_{\alpha\beta} \rangle_t + \langle T^m_{\alpha\beta} \rangle_t \right] \cdot ds$ | Optical Pressure |
|---|---|---|---|---|
| Drude | 151 | -849 | -698 | Negative |
| PEC | 487.3 | -299.5 | 187.9 | Positive |

**Supplementary material for the manuscript: "Strong light-induced negative optical pressure arising from the kinetic energy of conduction electrons in plasmonic cavities"**

H. Liu[1,2*], Jack Ng[2], S. B. Wang[2], Z. F. Lin[2, 3], Z. H. Hang[2], C. T. Chan[2,*] and S. N. Zhu[1]

[1]*Department of Physics, National Laboratory of Solid State Microstructures, Nanjing University, Nanjing 210093, People's Republic of China*

[2] *Department of Physics and Nano Science and Technology Program , Hong Kong University of Science and Technology, Clearwater Bay, Hong Kong, China*

[3]*Department of Physics, Fudan University, Shanghai, China*

**Corresponding Authors: liuhui@nju.edu.cn; phchan@ust.hk**

**Part I: Derivation of the values of various system parameters: magnetic inductance $L_m$, kinetic inductance $L_e$, capacitance $C_m$ and resistance $R_{eff}$**

**We first give the field patterns of the resonance mode in Fig. S.1 and an equivalent lumped element circuit for the patch-slab system in Fig. S.2.**

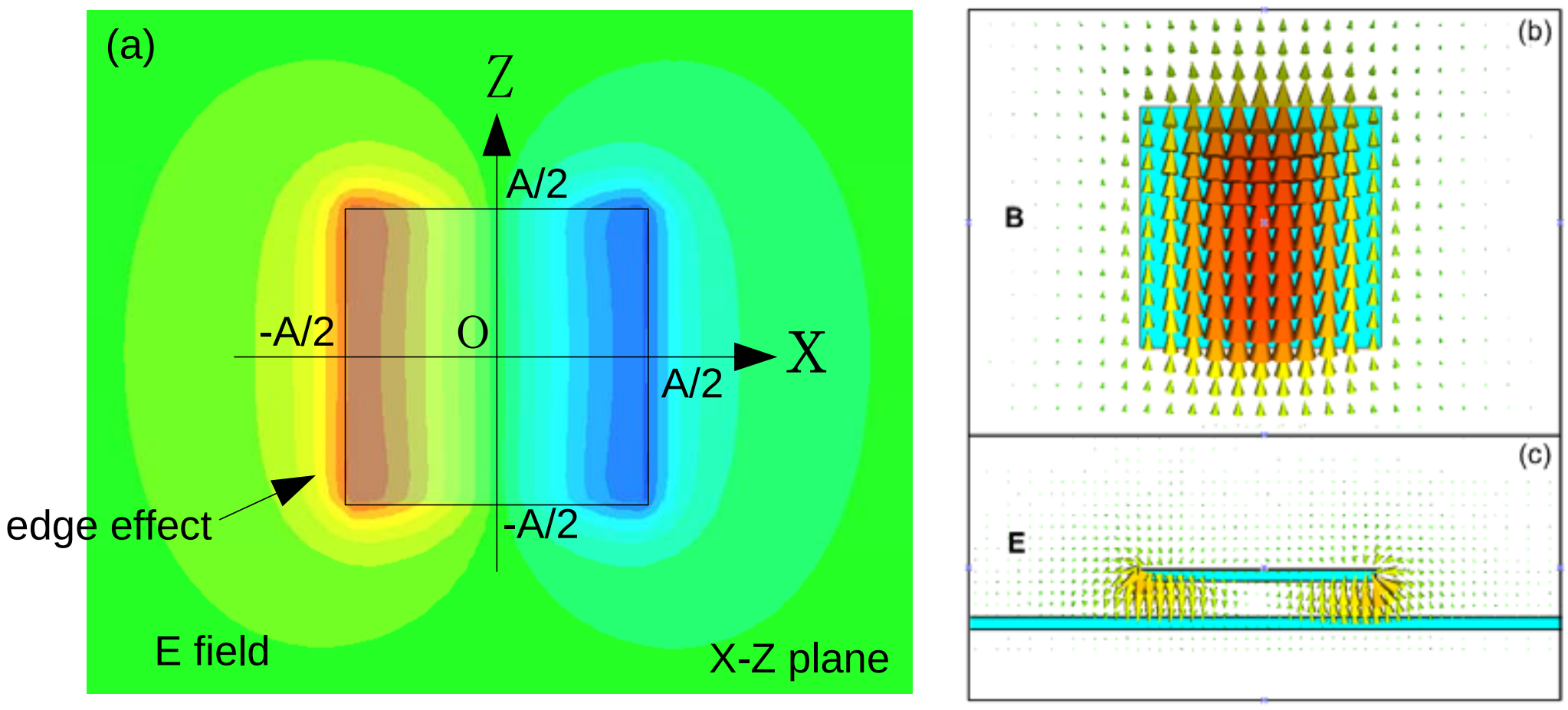

FIG. S.1 The profile of the electric field (panel a for y component and panel c) and of the magnetic field ( panel b) at the LC resonance (Note: By "edge effect", we mean that the electric field at the edge of patch is not entirely confined inside the space between patch and slab and spills outside –A/2 < X < A/2). Results are calculated with the numerical solver "CST".

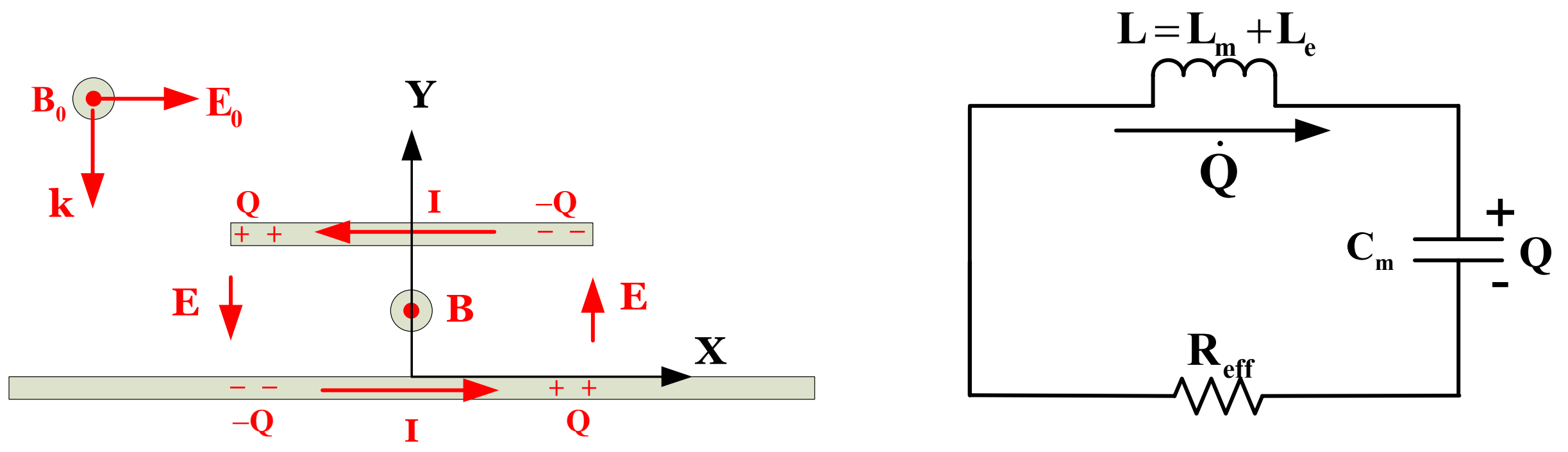

FIG S.2 Equivalent lumped LC circuit of patch-slab system

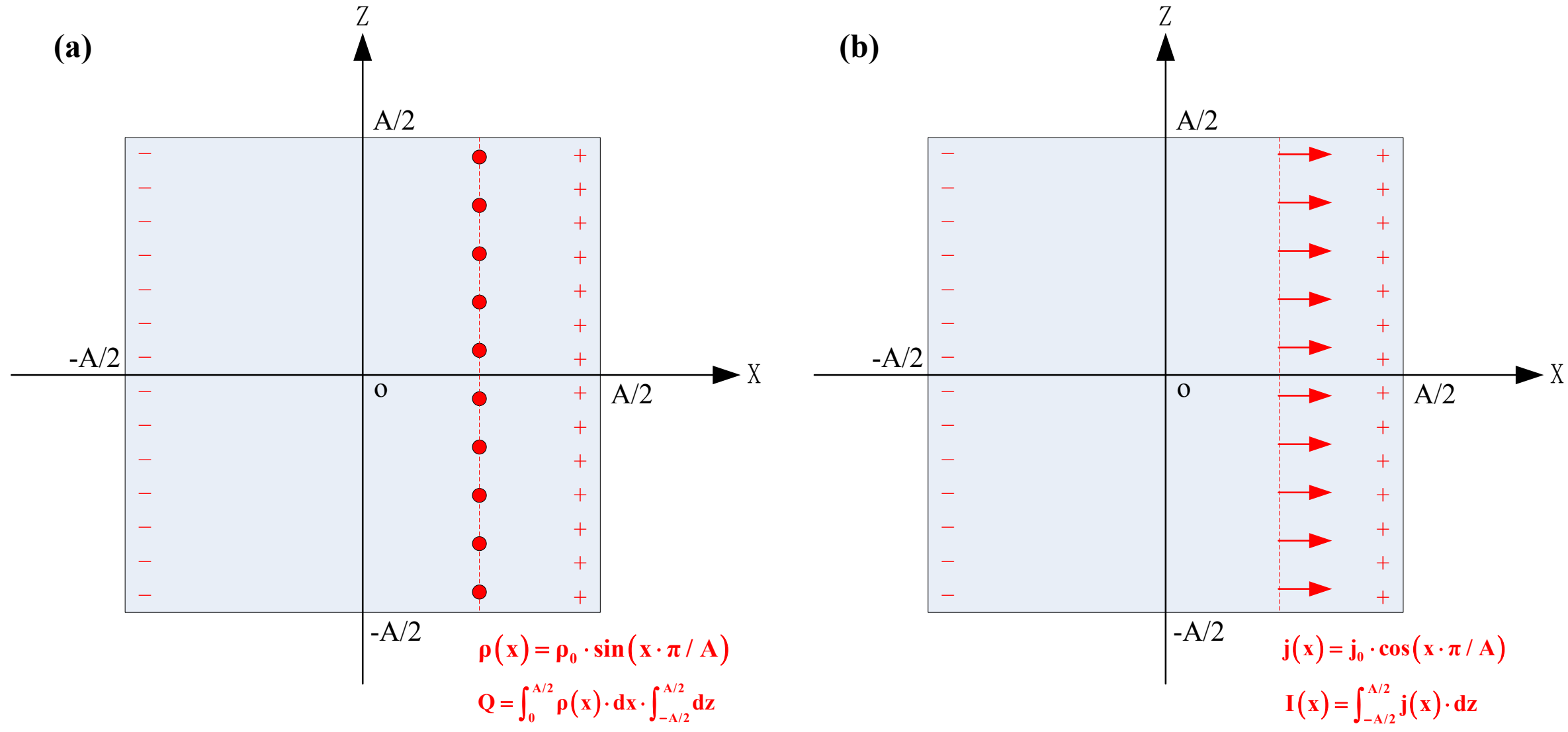

FIG S.3 (a) A schematic picture of the area density of charge $\rho(x)$ ; (b) Linear density of current $j(x)$ .

For the LC resonance mode depicted in Figure S.3, the external field induces a current to oscillate to and fro in the patch, leading to electric charge accumulation at two boundaries of patch during one oscillation cycle. By examining the numerically calculated field pattern, we found that for the fundamental resonance mode, it is a reasonably good appropriation to write the amplitude of the area density of charge on the patch as $\rho(x) = \rho_0 \cdot \sin(x \cdot \pi / A)$. Then, the total charge accumulated at one side of patch is $Q = \int_0^{A/2} \rho(x) \cdot dx \cdot \int_{-A/2}^{A/2} dz = \rho_0 \cdot A^2 / \pi$. At the same time, we can take the linear density of current on the patch is $j(x) = j_0 \cdot \cos(x \cdot \pi / A)$ , then the current passing through the patch at the location x is $I(x) = \int_{-A/2}^{A/2} j(x) \cdot dz = j_0 \cdot A \cdot \cos(x \cdot \pi / A)$. In the middle of patch, the current reach its maximum value $I_0 = j_0 \cdot A$ .

**I. Faraday magnetic inductance:**

The Faraday inductance $L_m$ is obtained from $E_m = \frac{1}{2} L_m I_0^2 \Rightarrow L_m = \frac{2E_m}{I_0^2}$ , where $E_m$ is the energy due to the magnetic field.

For the LC resonance mode, most of the magnetic energy is well confined inside the space between patch and slab. Magnetic field inside the gap can be approximated as $H(x)=j(x)$. Then the magnetic energy can be calculated as

$$E_m=\oiiint\left(\frac{\mu_0}{2}\cdot H(x)^2\right)\cdot dV=\frac{\mu_0}{2}\cdot\int_{-A/2}^{A/2}j(x)^2\cdot dx\cdot\int_{-A/2}^{A/2}dz\cdot\int_0^D dy=\frac{\mu_0}{4}\cdot D\cdot I_0^2 \qquad (1)$$

Here, magnetic inductance could be obtained as $L_m=2\cdot E_m/I_0^2=\frac{1}{2}\mu_0 D$.

## II. Kinetic or Internal inductance:

We use $E_k=\frac{1}{2}L_e I_0^2\Rightarrow L_e=\frac{2E_k}{I_0^2}$ to obtain the internal inductance where $E_k$ is the kinetic energy of the electrons in the induced currents inside the patch and the slab. For the LC resonance mode, part of the resonant mode energy is transformed into the kinetic energy of electrons inside the slab and patch when the field penetrates the metal. The velocity of electron is $v_e(x)=\frac{I(x)}{n_e\cdot e\cdot A\cdot t}=\frac{1}{n_e\cdot e\cdot t}\cdot j(x)$. The kinetic energy can be calculated as

$$E_e=\sum\left[\frac{1}{2}m_e\cdot v_e^2(x)\right]=\oiiint n_e\cdot\left[\frac{1}{2}m_e\cdot v_e^2(x)\right]\cdot dV=\frac{n_e m_e}{2}\cdot\frac{1}{n_e^2 e^2 t^2}\cdot\oiiint j(x)^2\cdot dV=\frac{1}{2}\cdot\frac{1}{\varepsilon_0\omega_p^2 t}\cdot I_0^2 \qquad (2)$$

Here, $\omega_p^2=\frac{n_e e^2}{\varepsilon_0 m_e}$. Then kinetic inductance could be obtained as $L_e=2E_e/I_0^2=\frac{1}{\varepsilon_0\omega_p^2 t}$. The result is good as long as the metal patch is thin so that it is fully penetrated by the external field. The integration volume includes the patch and the slab.

## III. Capacitance:

The capacitance can be obtained as $E_e=\frac{Q^2}{2C_m}\Rightarrow C_m=\frac{Q^2}{2E_e}$, where $E_e$ is the electric field energy in the cavity. If the electric field energy is well confined inside the space between patch and slab, it can be approximated as

$$E_e=\oiiint\left(\frac{1}{2}\varepsilon_0 E(x)^2\right)\cdot dV=\frac{1}{2}\varepsilon_0\oiiint\left[\frac{\rho(x)}{\varepsilon_0}\right]^2\cdot dV=\frac{1}{2\varepsilon_0}\int_{-A/2}^{A/2}\rho^2(x)dx\cdot\int_{-A/2}^{A/2}dz\cdot\int_0^D dy=\frac{1}{4\varepsilon_0}\rho_0^2\cdot A^2\cdot D=\frac{\pi^2}{4\varepsilon_0}\frac{Q^2}{A^2}\cdot D \qquad (3)$$

Then the capacitance is $C_m=\frac{Q^2}{2E_e}=\varepsilon_0\frac{2}{\pi^2}\frac{A^2}{D}$. Actually, for the LC resonance mode given in figure S.3, we can see that electric field energy is not that well confined between the patch and slab due to the edge effect at the boundary of patch. It is difficult to calculate the total electric energy of the whole system analytically including the edge effect. However, the calculation of total capacitance should include the edge effect. Here, we could introduce an empirical coefficient $\alpha$ to denote the magnitude of edge effect and the total capacitance could be expressed as $C_m=\varepsilon_0\frac{2}{\pi^2}\frac{A^2}{D}\cdot\left(1+\alpha\frac{D}{A}\right)$. Here, the coefficient $\alpha$ could be obtained from simulation data (see Figure S.4).

## IV. Effective Resistance

The complex power of this LC resonator is obtained by $P=\frac{1}{2}Z_{eff}\cdot I_0^2$ where Z is the effective impedance of the system. For the whole system, the complex power can be calculated by

$$P = 2\int_{-A/2}^{A/2}\left[\frac{1}{2}\cdot I(x)^2\cdot dZ(x)\right] = 2\int_{-A/2}^{A/2}\left[\frac{1}{2}\cdot j(x)^2\cdot A^2\cdot\frac{1}{\sigma}\cdot\frac{dx}{A\cdot t}\right] = \frac{1}{2}\cdot\frac{1}{\sigma\cdot t}\cdot I_0^2 . \quad (4)$$

The factor of 2 takes care of the patch and the slab.

Here, the complex conductivity is $\sigma = i\varepsilon_0\frac{\omega_p^2}{\omega + i\gamma_m}$ for drude model. Then the total effective impedance is

$$Z_{eff} = \frac{1}{\sigma}\cdot\frac{1}{t} = \frac{\gamma_m}{\varepsilon_0\omega_p^2 t} - i\frac{\omega}{\varepsilon_0\omega_p^2 t} = R_{eff} - i\omega L_e . \quad (5)$$

Then we obtain the Ohmic resistance as $R_{eff} = \frac{\gamma_m}{\varepsilon_0\omega_p^2 t}$ and $L_e = \frac{1}{\varepsilon_0\omega_p^2 t}$ (this result for $L_e$ agrees the above calculation for $L_e$).

## V. Comparison with simulation data

The patch–slab structure can be seen as an equivalent LC circuit. Based on the above calculation of $L_m$, $L_e$ and $C_m$, its resonance frequency could be written as

$$\omega = \frac{1}{\sqrt{(L_e + L_m)\cdot C_m}} = \frac{1}{\sqrt{\left(\frac{1}{\varepsilon_0\omega_p^2 t} + \frac{\mu_0 D}{2}\right)\cdot\varepsilon_0\frac{2}{\pi^2}\frac{A^2}{D}\cdot\left(1+\alpha\cdot\frac{D}{A}\right)}} . \quad (6)$$

In the simulations, the thickness is set as t=10nm. We could obtain the dependence of resonance frequency on A and D, see figure 4. Fitting the simulated data with equation (6), we could obtain a value of $\alpha = 14$ for the best fit.

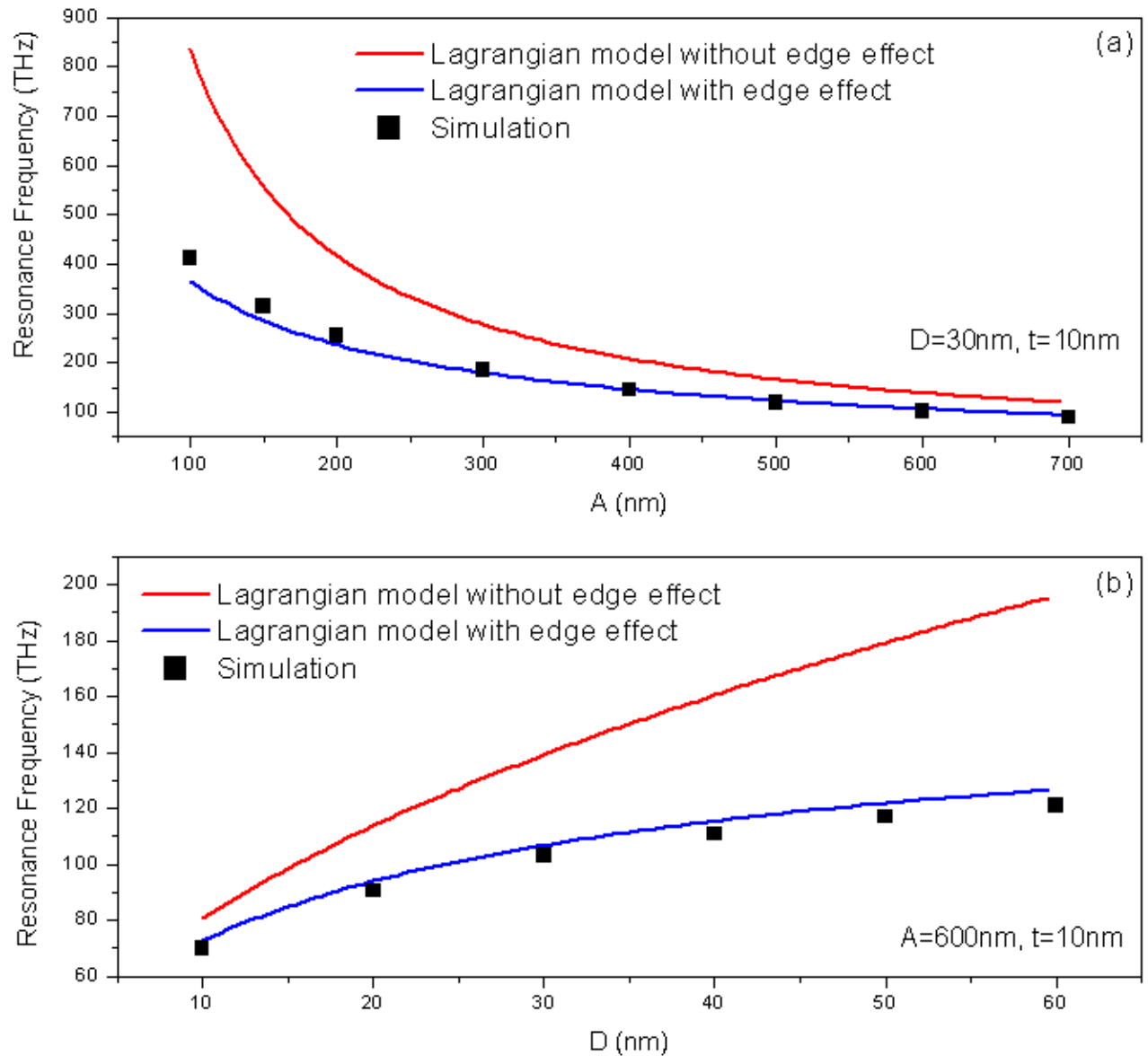

FIG. S.4 The dependence of resonance frequency on A (with D=30nm) and D (with A=600nm). Here, black dots: simulated data; red solid line: calculated results from equation (6) without edge effect ($\alpha = 0$); blue solid line: calculated results from equation (6) including edge effect ($\alpha = 14$)

**Part II: Derivation of optical pressure from Lagrangian model**

In Fig. S.1-3, the system of a patch and a slab can be seen as an equivalent LC circuit resonator. Here, we propose a Lagrangian model to describe the resonance property of the equivalent LC circuit for the resonator. The circuit's Lagrangian can be expressed as

$$\mathcal{L} = \frac{L\cdot\dot{Q}^2}{2} - \frac{Q^2}{2\cdot C}\,. \tag{7}$$

Here, $L = L_m + L_e$ and $C = C_m + C_e$. $L_m = \frac{1}{2}\mu_0 D$ and $C_m = \varepsilon_0 \frac{2}{\pi^2}\frac{A^2}{D}\cdot\left(1+\alpha\frac{D}{A}\right)$ are the external inductance and capacitance, $L_e = \frac{1}{\omega_p^2\varepsilon_0 t}$ and $C_e$ are the internal inductance and capacitance. We will ignore the internal capacitance $C_e$ which is much smaller than external capacitance $C_m$. Q is the net charge in the capacitor, $L_m\dot{Q}^2/2$ is the magnetic field energy stored between the two patches, $U_{kinetic} = L_e\dot{Q}^2/2$ is the kinetic energy of the electrons due to the induced current inside the metal. We will ignore the internal capacitance which is much smaller than external capacitor, i.e. $C \approx C_m$. Then $U_{electric} = Q^2/(2C_m)$ is the potential energy stored in the cavity.

In the presence of Ohmic dissipation and an external driving field, the Euler-Lagrange equation can be written as $\frac{d}{dt}\left(\frac{\partial\mathcal{L}}{\partial\dot{Q}}\right) - \frac{\partial\mathcal{L}}{\partial Q} = -\frac{\partial\mathcal{R}}{\partial\dot{Q}} + \text{e.m.f.}$, where $\mathcal{R} = R_{eff}\cdot\dot{Q}^2/2$ ($R_{eff}$ is the effective resistance in the system) and the e.m.f. due to external field $E_0$ is $\text{emf} = -\oint E\cdot d\ell = -\left(\ell_{eff}^{(1)}\cdot E_0 - \ell_{eff}^{(2)}\cdot E_0\cdot e^{ikD}\right)$, with $\ell_{eff}^{(1)}$ and $\ell_{eff}^{(2)}$ being the effective length of the dipole on the patch and the slab respectively and the factor $\exp(ik\cdot D) = \exp\left(i\frac{\omega}{c}\cdot D\right)$ represents the retardation effect. To simplify expressions, we introduce the notation $\ell_{eff}^{'} = \ell_{eff}^{(1)}\cdot\left(1-\tau\cdot e^{ikD}\right)$ with $\tau = \ell_{eff}^{(2)}/\ell_{eff}^{(1)}$. Solving the Euler-Lagrange equation gives the resonance frequency of the cavity $\omega_0 = \frac{1}{\sqrt{(L_e+L_m)\cdot C_m}} = \frac{1}{\sqrt{\left(\frac{1}{\varepsilon_0\omega_p^2 t}+\frac{\mu_0 D}{2}\right)\cdot\varepsilon_0\frac{2}{\pi^2}\frac{A^2}{D}\cdot\left(1+\alpha\cdot\frac{D}{A}\right)}}$. We obtain the solution for Q as

$$Q = -\frac{\ell_{eff}^{'}}{L}\cdot\frac{1}{\sqrt{\left(\omega^2-\frac{1}{LC}\right)^2+\left(\frac{R_{eff}}{L}\omega\right)^2}}\cdot E_0\,. \tag{8}$$

At the resonance $\omega_0 = 1/\sqrt{LC}$, the solution of charge is $Q(\omega=\omega_0) = -\frac{\ell_{eff}^{'}}{R_{eff}}\cdot\frac{1}{\omega_0}\cdot E_0$. The optical force exerted on the patch can be calculated as a generalized force corresponding to the generalized coordinate D

$$\mathcal{F} = \frac{\partial\mathcal{L}}{\partial D} = \frac{\partial}{\partial D}\left(\frac{L\dot{Q}^2}{2} - \frac{Q^2}{2C_m}\right) = \frac{1}{2}\left(\frac{\partial L_m}{\partial D}+\frac{\partial L_e}{\partial D}\right)\dot{Q}^2 + \frac{1}{2}\frac{Q^2}{C_m^2}\left(\frac{\partial C_m}{\partial D}\right). \tag{9}$$

If we include E field edge effect, the capacitor could be expressed as $C_m = \varepsilon_0\frac{2}{\pi^2}\frac{A^2}{D}\cdot\left(1+\alpha\frac{D}{A}\right)$ (see Part I), then we have $\frac{\partial C_m}{\partial D} = -\frac{C_m}{D}\cdot\frac{A}{A+\alpha\cdot D}$. As the internal inductance does not depend on the gap distance, we have $\partial L_e/\partial D = 0$. For magnetic inductance, we have $L_m = \frac{1}{2}\mu_0 D \Rightarrow \partial L_m/\partial D = L_m/D$. Then time average of optical force could be expressed as

$$
\begin{aligned}
\langle \mathcal{F} \rangle_t &= \frac{1}{2}\frac{\partial L_m}{\partial D}\cdot\langle \dot{Q}^2\rangle_t + \frac{1}{2}\frac{\langle Q^2\rangle_t}{C_m^2}\cdot\frac{\partial C_m}{\partial D} \\
&= \frac{1}{2}\frac{L_m}{D}\langle \dot{Q}^2\rangle_t - \frac{1}{2}\frac{\langle Q^2\rangle_t}{C_m}\frac{1}{D}\left(\frac{A}{A+\alpha\cdot D}\right) \\
&= \left(\frac{1}{2}\frac{L_m}{D}\langle \dot{Q}^2\rangle_t - \frac{1}{2}\frac{\langle Q^2\rangle_t}{C_m}\frac{1}{D}\right) + \frac{1}{2}\frac{\langle Q^2\rangle_t}{C_m}\frac{1}{D}\left(\frac{\alpha D}{A+\alpha D}\right) \quad \left(\text{note}: \omega_0^2 = \frac{1}{\sqrt{LC_m}};\ \langle \dot{Q}^2\rangle_t = \omega_0^2\langle Q^2\rangle_t\right) \\
&= \left(\frac{1}{2}\frac{L_m}{D}\langle \dot{Q}^2\rangle_t - \frac{1}{2}\frac{L}{D}\langle \dot{Q}^2\rangle_t\right) + \frac{1}{2}\frac{\langle Q^2\rangle_t}{C_m}\frac{1}{D}\left(\frac{\alpha D}{A+\alpha D}\right) \\
&= -\frac{1}{2}\frac{L_e\langle \dot{Q}^2\rangle_t}{D} + \frac{1}{2}\frac{\langle Q^2\rangle_t}{C_m}\frac{1}{D}\left(\frac{\alpha D}{A+\alpha D}\right).
\end{aligned} \tag{10}
$$

Here, we can see that the total optical force given in equation (10) includes two parts: $-\frac{1}{2}\frac{L_e\langle \dot{Q}^2\rangle_t}{D}$ is an attractive force originating from kinetic energy of electron; $\frac{1}{2}\frac{\langle Q^2\rangle_t}{C_m}\frac{1}{D}\left(\frac{\alpha D}{A+\alpha D}\right)$ is a repulsive force coming from E field edge effect. At resonance frequency $\omega=\omega_0$, $\frac{1}{2}\frac{\langle Q^2\rangle_t}{C_m}=\frac{1}{2}\frac{(L_m+L_e)\langle \dot{Q}^2\rangle_t}{D}$ and $\langle \dot{Q}^2\rangle_t=\left(\frac{\ell_{eff}^{(1)}}{R_{eff}}\right)^2\cdot\left[\frac{1+\tau^2}{2}-\tau\cdot\cos\left(\frac{\omega_0}{c}\cdot D\right)\right]\cdot E_0^2$. The effective length of patch could be approximately taken as $\ell_{eff}^{(1)}\simeq A$. In our design, the patch is square shape. Then the area of patch is $S=A^2\simeq\left(\ell_{eff}^{(1)}\right)^2$. The optical pressure normalized by the incident intensity $I_0=\frac{1}{2}\varepsilon_0\cdot c\cdot E_0^2$ could be obtained as

$$
P=\frac{\langle \mathcal{F}\rangle_t}{I_0\cdot S^2}=\frac{L_e}{\varepsilon_0 c R_{eff}^2}\frac{1}{D}\cdot\left[\frac{1+\tau^2}{2}-\tau\cdot\cos\left(\frac{\omega_0}{c}\cdot D\right)\right]\cdot\left(-1+\frac{L}{L_e}\cdot\frac{\alpha D}{A+\alpha D}\right). \tag{11}
$$

In equation (11), $\omega_0=1/\sqrt{L\cdot C_m}$, $L=L_m+L_e$, $L_e=\frac{1}{\omega_p^2\varepsilon_0}\cdot\frac{1}{t}$, $L_m=\frac{1}{2}\mu_0 D$, $C_m=\varepsilon_0\frac{2}{\pi^2}\frac{A^2}{D}\cdot\left(1+\alpha\frac{D}{A}\right)$, $R_{eff}=\frac{\gamma_m}{\omega_p^2\varepsilon_0}\cdot\frac{1}{t}$. (for gold, $\omega_p=1.37\times10^{16}\,Hz$, $\gamma_m=12.08\times10^{13}\,Hz$ ). At last, we cold obtain the normalized optical pressure as

$$
P=\frac{\eta}{D}\cdot\left[\frac{1+\tau^2}{2}-\tau\cdot\cos\left(2\pi\sqrt{\frac{\beta D}{(1+\beta D)}\cdot\frac{\pi^2\cdot D^2}{A(A+\alpha D)}}\right)\right]\cdot\left(-1+(1+\beta D)\cdot\frac{\alpha D}{A+\alpha D}\right). \tag{12}
$$

Where $\eta=\frac{t}{c}\cdot\frac{\omega_p^2}{\gamma_m^2}$, $\beta=\frac{t}{2}\cdot\frac{\omega_p^2}{c^2}$. We could find that the normalized optical pressure P depends on A and D. In equation (12), $\alpha$ comes from edge effect of capacitor, $\tau$ come from the ratio between effective lengths of patch and slab which are generally different. For t=10nm, based on the resonance frequency data given in figure 3, we could obtain a value of $\alpha=14$ (see Part I). In Fig.5, we compare the calculated optical pressure from equation (12) and the simulated data. We found when $\tau=0.77$, Lagrangian model agree with simulation quite well.

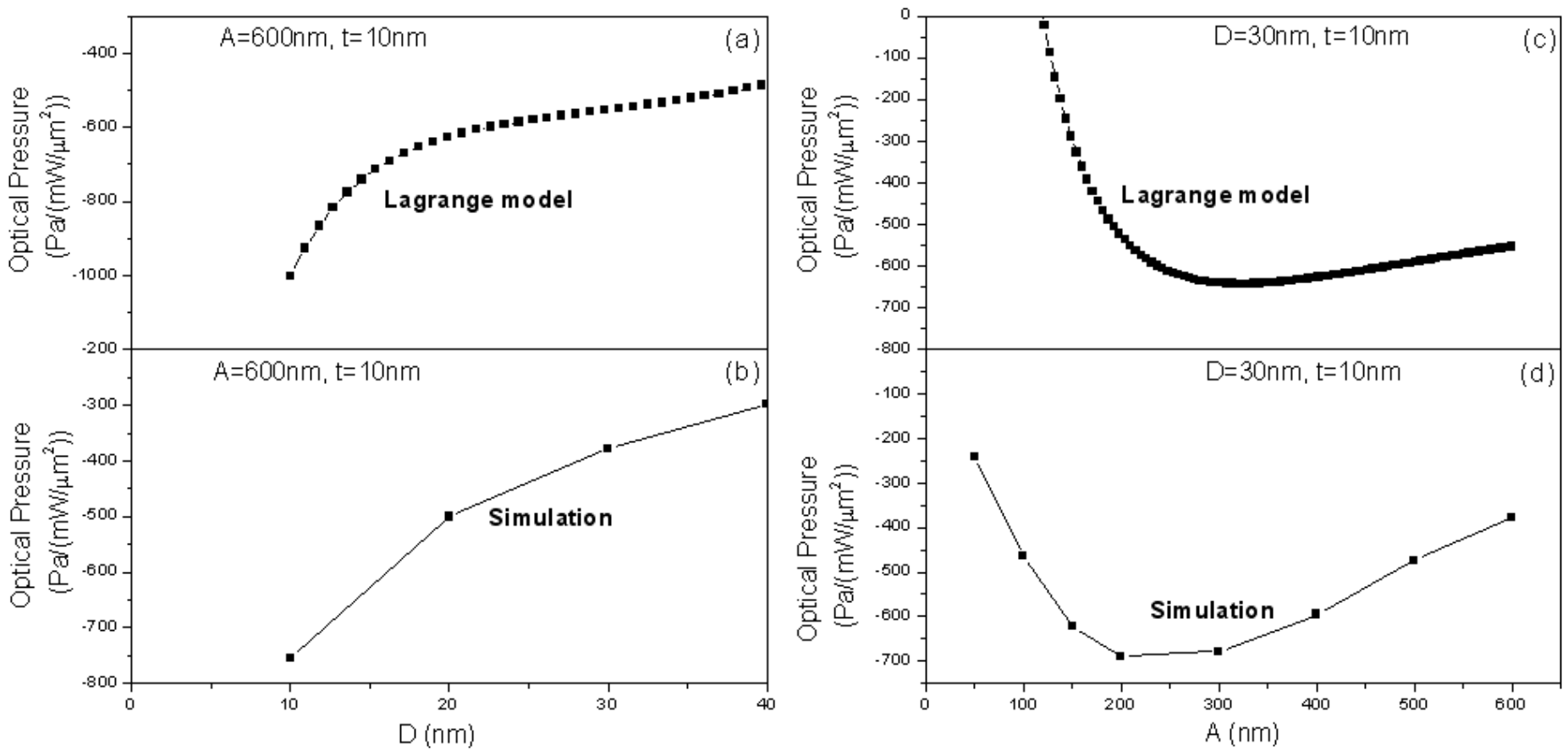

FIG. S.5 Comparison between the Lagrangian model and simulation results: (a) D dependence of optical pressure calculated from Lagrangian model, which shows a bigger force for a small D; (b) D dependence of optical pressure from simulated data; (c) A dependence of optical pressure calculated from Lagrangian model; (b) A dependence of optical pressure from simulated data. Note that there is an optimal patch size that gives the maximal attractive force.